\documentclass[a4paper,11pt]{article}
\usepackage{pos}
\usepackage{hyperref}
\usepackage{caption}
\usepackage{subcaption}

\title{Design optimization of hadronic calorimeters for future colliders}
\author*[a,b]{Bruno Rodrigues}
\author[b]{Inês Ochoa}
\author[a,b]{Agostinho Gomes}

\affiliation[a]{Faculty of Sciences of the University of Lisbon,\\
 1749-016, Lisboa, Portugal}

\affiliation[b]{LIP - Laboratory of Instrumentation and Experimental Particle Physics,\\
 1649-003, Lisboa, Portugal}

\emailAdd{brodrigues@lip.pt}
\emailAdd{miochoa@lip.pt}

\abstract{
Calorimeters are a crucial component in modern particle detectors. They are responsible for providing accurate energy measurements of particles produced in high-energy collisions. The demanding requirements set for next-generation collider experiments impose new challenges on the design of new detectors, and a systematic approach to their optimization is increasingly necessary. 
The performance of calorimeters is primarily characterized by their energy resolution, parameterized by a stochastic and a constant term, related to sampling fluctuations and non-uniformities respectively. To improve the reconstruction quality of physics objects in the calorimeter, both terms need to be taken into account. Changes in a longitudinally constrained design usually result in a trade-off between these terms, making optimization a non-trivial task.
This work focuses on the optimization of a hadronic sampling calorimeter, based on the FCC-ee ALLEGRO detector concept. By controlling the absorber layer thickness in a Geant4 simulation, the impact of the passive to active material proportion on the deposited energy distribution and resolution can be analyzed.  
Our methodology aims at exploring the design space with practical considerations, paving the way for the development of a closed optimization framework that can evaluate multiple designs against physics performance targets.
}

\FullConference{Fifth MODE Workshop on Differentiable Programming for Experiment Design (MODE2025)\\
8-13 June 2025\\
Kolymbari, Crete, Greece\\}

\begin{document}
\maketitle

\section{Introduction}

Sampling calorimeters use passive and active materials to contain and measure the energy deposited by incoming particles. The passive absorber materials are chosen to have a shorter interaction length ($\lambda_{int}$) and quickly initiate showers of secondary particles when traversed by energetic particles produced in collisions. As the showers propagate, their energy is deposited in the active material, in this case scintillators, which produce light that is then read out by sensitive detectors, such as photomultipliers. Hadron showers have electromagnetic and hadronic components (em and non-em) categorized by the energy fraction in the em component ($f_{em}$) and in the non-em component ($1-f_{em}$). The energy of hadrons is especially difficult to measure accurately because a considerable amount of the hadronic component is lost in nuclear interactions, with large event-to-event fluctuations. The response to electrons ($e$) and hadrons ($\pi$) is usually not the same, and the calorimeter is said to be non-compensating. It is suggested in \cite{fabiola} that achieving compensation, $e/\pi = 1$, is the key performance parameter of energy resolution, and by properly choosing the type and thickness of active and passive materials, the response can be tuned to obtain $e/\pi \approx 1$.

The energy resolution of a calorimeter represents its ability to identify and separate the signal of particles with different energies. For particles with energy $E$, it is defined as the standard deviation of the measured energies ($\sigma_E$), normalized by the mean: $\sigma_E/E$. The energy dependence can be modeled with three parameters added in quadrature using Equation \ref{eq:enery_res}.

\begin{equation}
    \frac{\sigma_E}{E} = \frac{a}{\sqrt{E}} \oplus \frac{b}{E} \oplus c.
    \label{eq:enery_res}
\end{equation}

The stochastic term, $a$, represents the intrinsic statistical fluctuations associated with the development of the shower, along with sampling fluctuations. The noise term, $b$, is due to the electronics in the readout chain. The constant term, $c$, arises from imperfections in the detector and other non-uniformities. Non-compensation and fluctuations in $f_{em}$ contribute to the energy resolution through an approximately constant term \cite{wigmans}, included in $c$.

\begin{figure}[hb]
    \centering
    \includegraphics[width=0.5\linewidth]{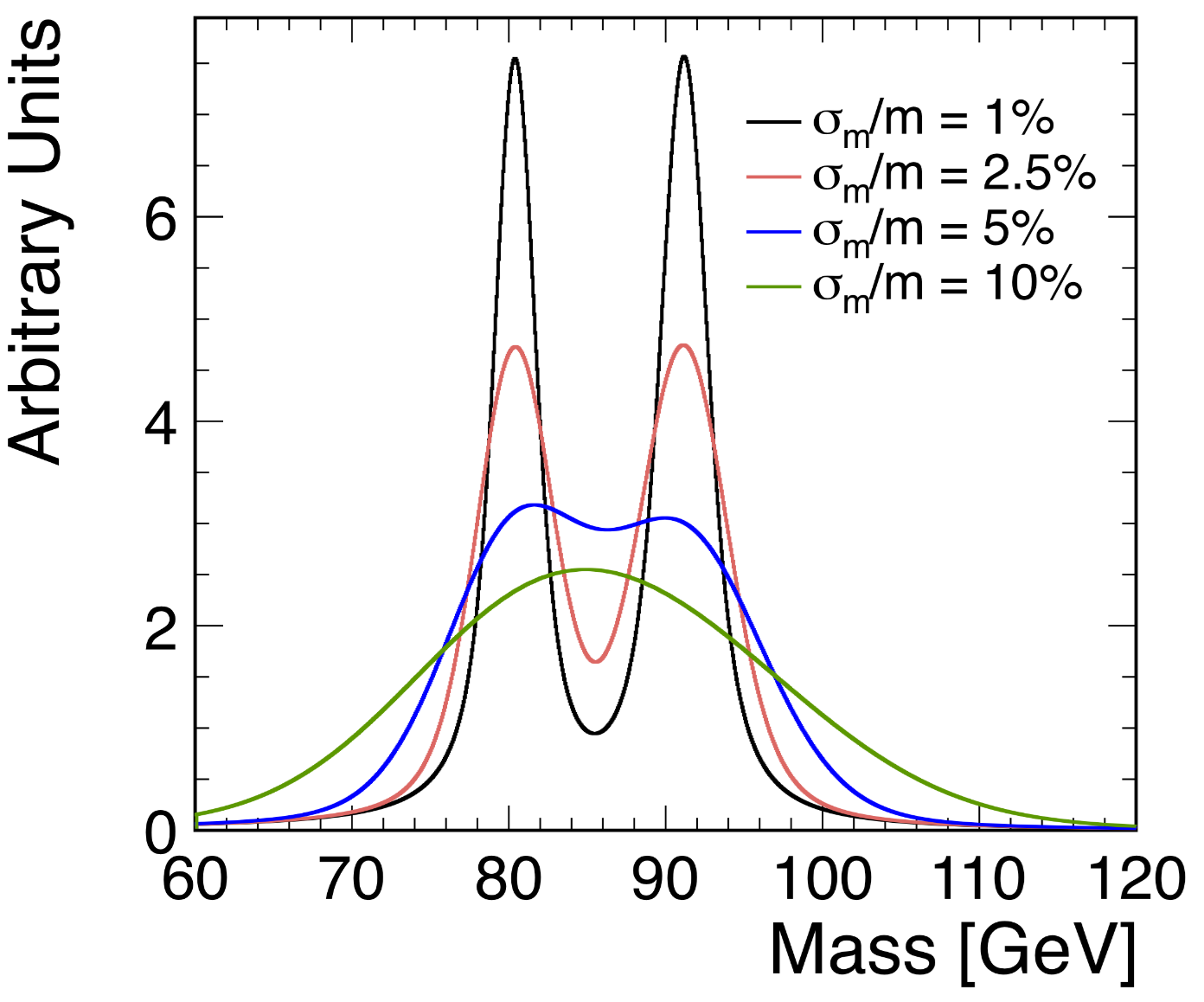}
    \caption{Impact of jet energy resolution on the separation of $W/Z$ invariant mass peaks. Image from \cite{seminar}.}
    \label{fig:W/Z}
\end{figure}

The separation of $W$ and $Z$ bosons is a requirement of the FCC-ee physics program \cite{fcc-ee-feasibility} and is driven by hadronic calorimeter resolution, illustrated in \ref{fig:W/Z}.

\section{Calorimeter simulation}

The Geant4 toolkit is used to accurately simulate the response of a calorimeter to particles produced in high-energy collisions. This toolkit is specialized in the simulation of the passage of particles through matter, allowing the user to define the geometry of the system, the materials involved, the fundamental particles of interest, the physics processes governing particle interactions, and the response of sensitive detector components. The Geant4 documentation \cite{geant4-doc} recommends the FTFP-BERT physics list for collider physics applications, as it usually produces the best agreement with test beam calorimeter data, including shower shape, energy response and resolution.

\begin{figure}[hb]
    \centering
    \includegraphics[width=0.85\linewidth]{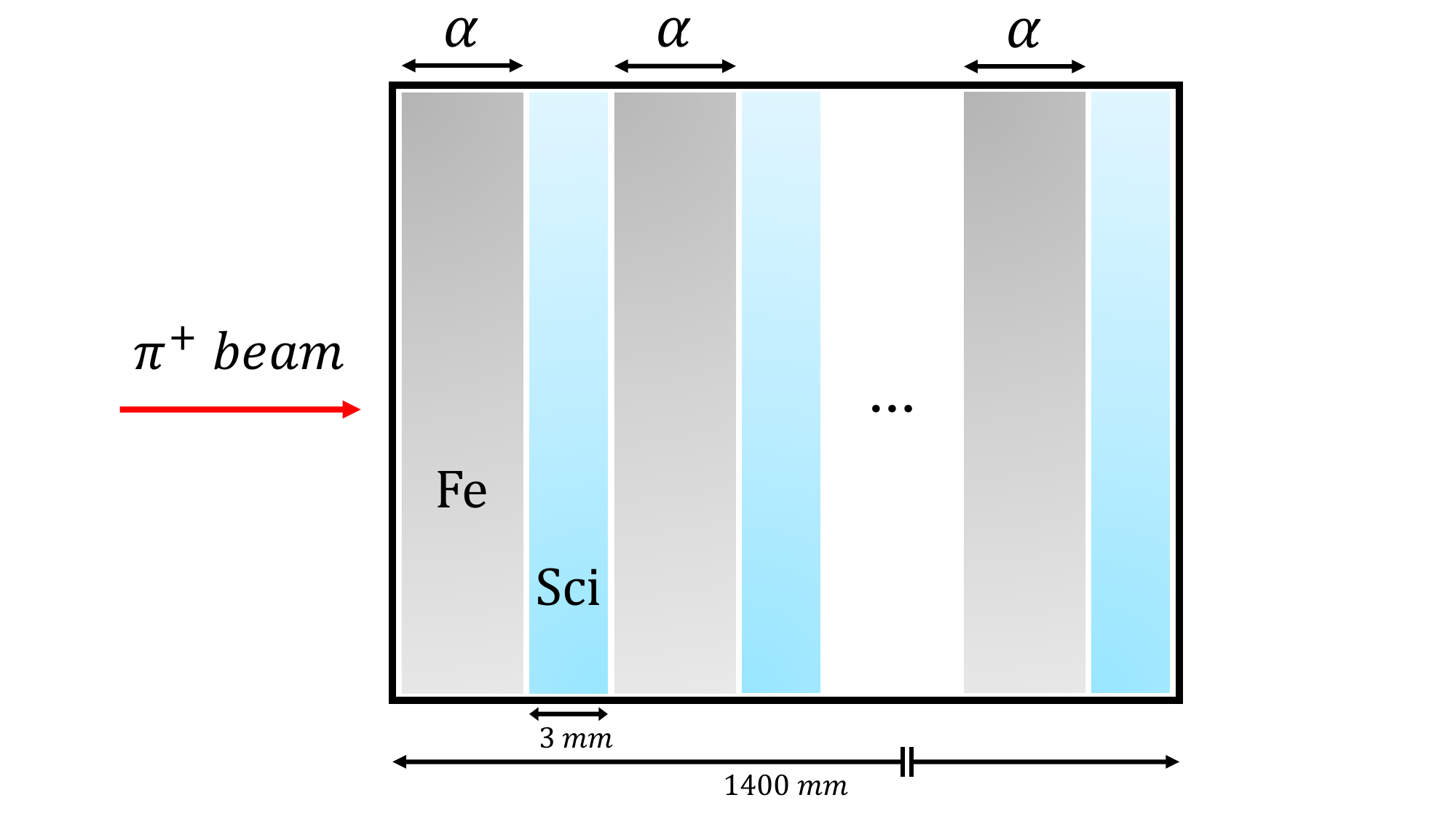}
    \caption{Schematic representation of the simulated calorimeter. The charged pions are shot in the longitudinal direction and hit the center of the square face of the calorimeter perpendicularly. The absorber plates are made of iron (Fe) and their thickness ($\alpha$) is variable. The plastic scintillator plates are composed of Polyvinyl toluene, with a thickness of 3~mm. The total depth of the calorimeter is fixed at 1400~mm.}
    \label{fig:calo_diagram}
\end{figure}

The cylindrical geometry of the hadronic barrel in \cite{allegro} can be approximated, for the purpose of this simulation, with a box of depth corresponding to the radial dimension of the barrel, and sides sufficiently large to contain lateral leakage. 
The simulated geometry has a 1000~mm wide square face that is first hit by the particles in the normal direction, they then traverse 1400~mm of alternating absorber and scintillator plates in depth, as illustrated in Figure \ref{fig:calo_diagram}. 

Changing the absorber thickness ($\alpha$) while constraining the total depth of the calorimeter to 1400~mm allows for a fair comparison of different absorber to scintillator proportions on the energy resolution, while maintaining the design within the specifications. Simply changing the thickness of the absorber would result in a deeper calorimeter, reducing potential leakage and unfairly benefiting thicker absorbers in the comparison. Thus, the number of absorber-scintillator plates changes dynamically with the selected thickness to ensure that the total depth is maintained constant. 

The single hadron response is simulated with a particle gun. Firing 5000 charged pions at eight different energy values ranging from 1~GeV to 200~GeV. The energy deposited in all the 3~mm thick plastic scintillators is then summed and measured as the detector response for each event. 

\section{Exploring the parameter space}

\begin{figure}[ht]
    \centering
    \includegraphics[width=0.85\linewidth]{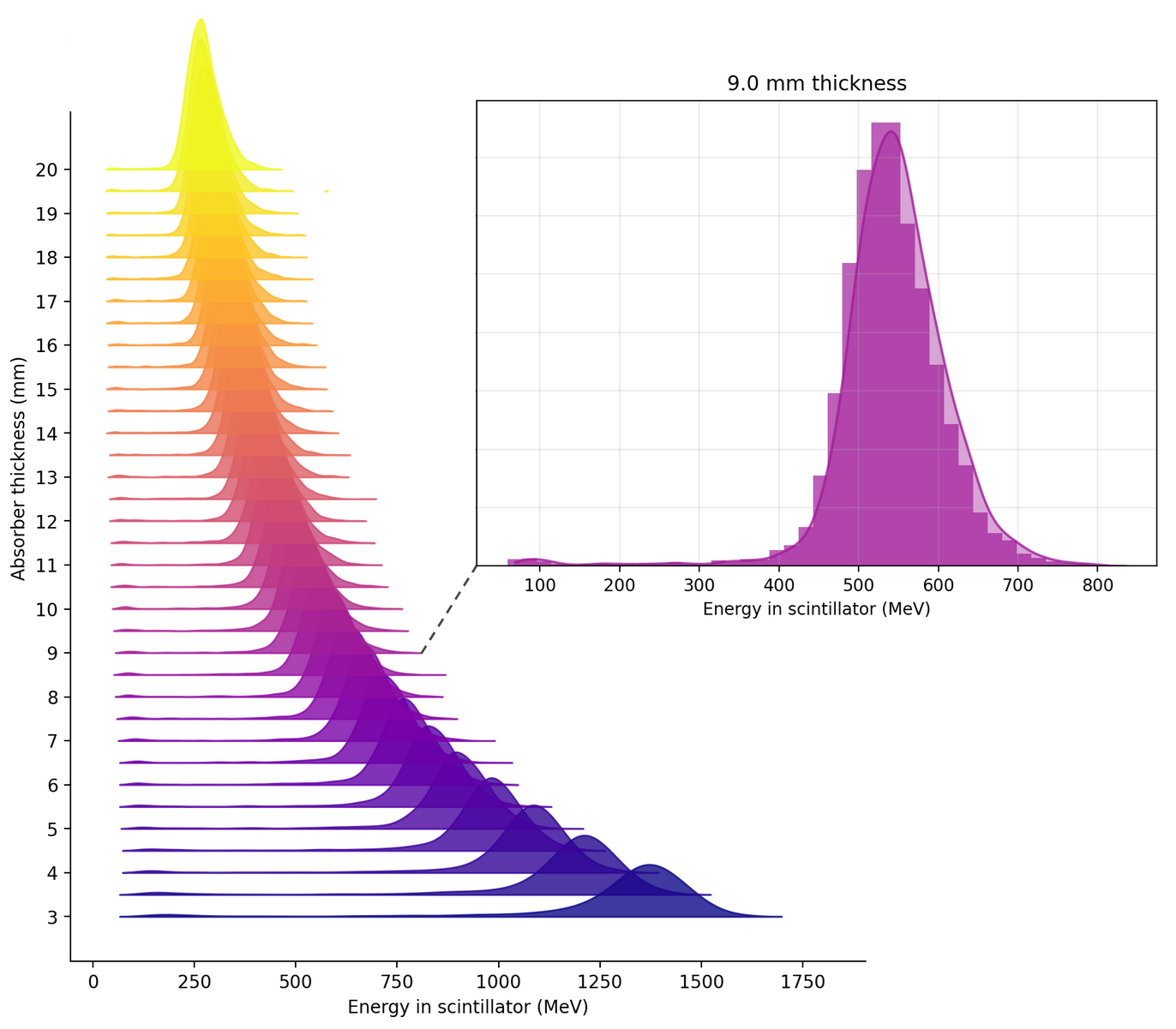}
    \caption{Energy deposited in the scintillator by 10~GeV $\pi^+$ for different absorber thicknesses. The histogram bins are shown in the expanded plot for a 9 mm thick absorber, along with a kernel density estimation (KDE) curve. For visual clarity, only estimates above $10^{-5}$ density are shown on the plot. The non-Gaussian shape of the distributions is an indicator of a non-compensating calorimeter. The small low energy peaks are excluded with sigma clipping and therefore don't directly impact this calculation of the resolution.}
    \label{fig:joyplot}
\end{figure}

To visualize the magnitude of event-to-event fluctuations and their distribution we can use a histogram of the deposited energy in the scintillator. The effect on the response caused by the change in absorber thickness for 10~GeV pions can be seen in Figure \ref{fig:joyplot}. A thicker absorber (larger absorber-to-scintillator proportion) reduces the sampling fraction, and thus the average energy deposited in the scintillator decreases; but it also helps to narrow the width of the distribution to a certain degree. Since the energy resolution is defined as $\sigma/E$, decreasing both $\sigma$ and $E$ does not necessarily result in an improvement. The goal of optimization is to find the ideal thickness that minimizes this quantity for all energies using the resolution parameters as a figure of merit. 

\begin{figure}[htb]
     \centering
     \begin{subfigure}[b]{0.49\linewidth}
         \centering
         \includegraphics[width=\linewidth]{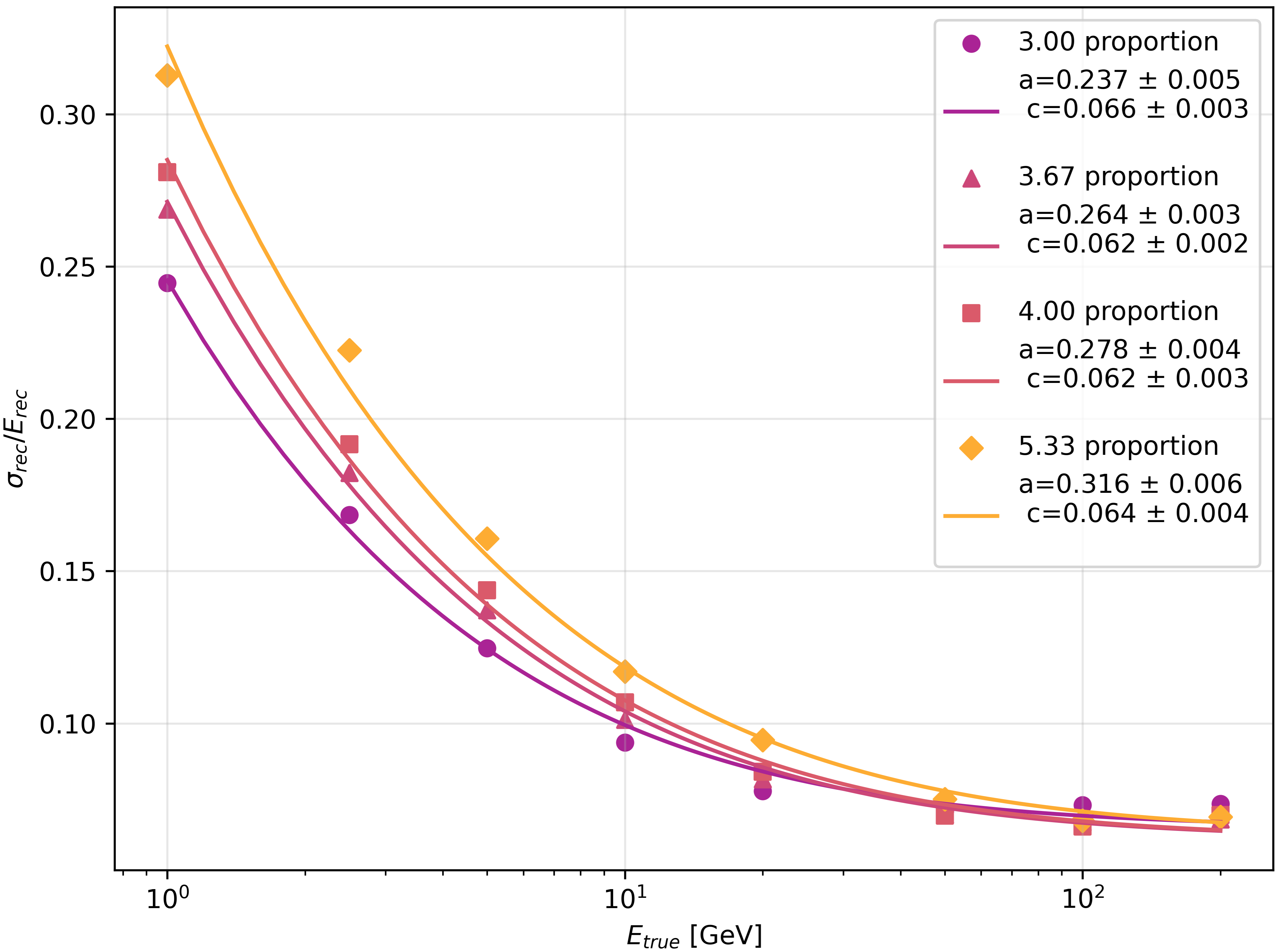}
     \end{subfigure}
     \hfill
     \begin{subfigure}[b]{0.49\linewidth}
         \centering
         \includegraphics[width=\linewidth]{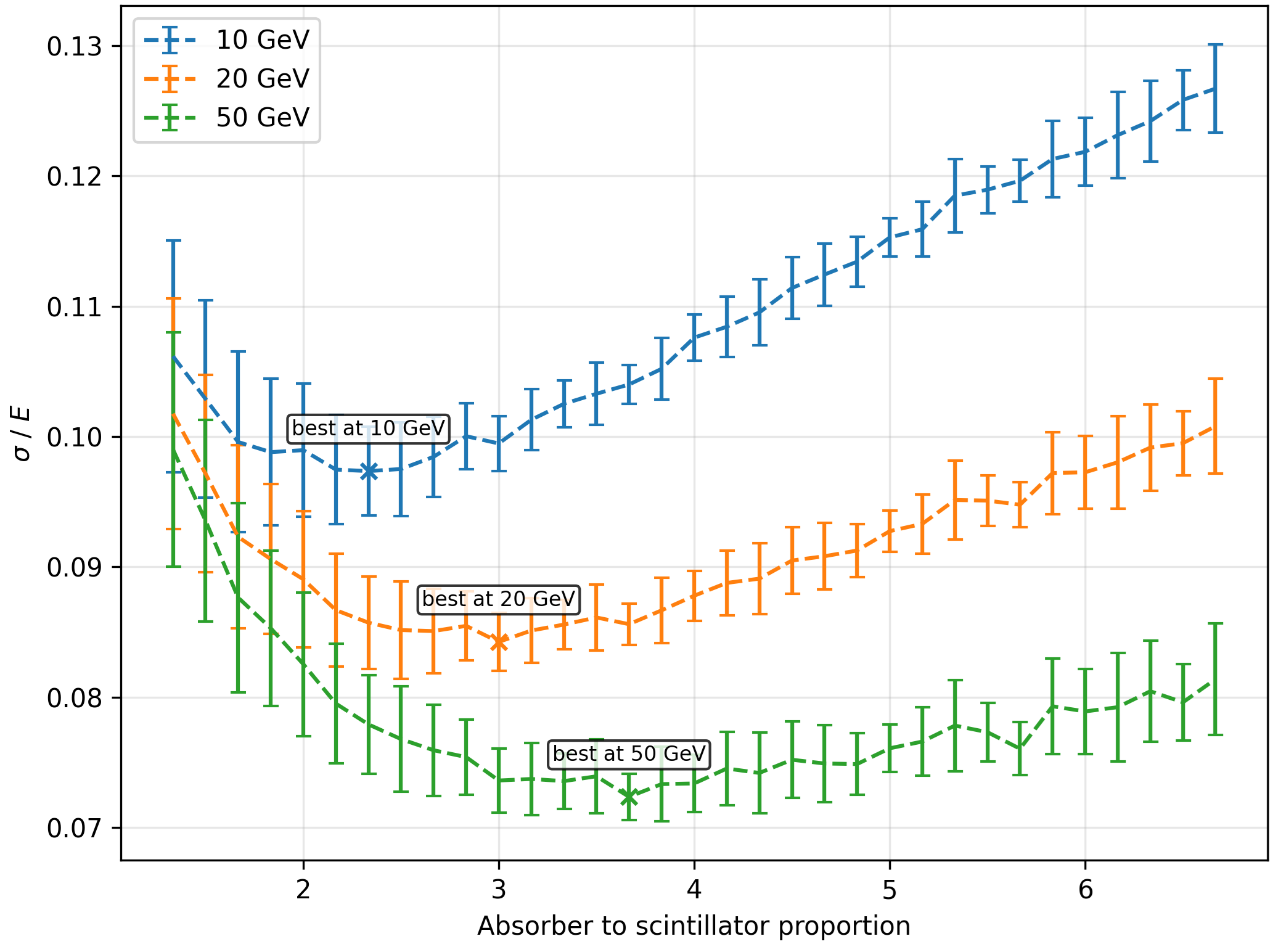}
     \end{subfigure}
     \hfill
        \caption{Left: Resolution curves for different absorber proportions. The response to charged pions ($\pi^+$) at the simulated energies is used to estimate the stochastic and constant parameters of the energy resolution by fitting Eq. \ref{eq:enery_res} to the points. Four selected proportions are shown. Right: Energy resolution as a function of absorber proportion for $\pi^+$ with three different energies. The best resolution is achieved with different proportions at different energies.}
        \label{fig:res_curve_prop}
\end{figure}

To calculate the energy resolution $\sigma_E/E$ for the different absorber proportions, we must first obtain the average $E$ and standard deviation $\sigma_E$ of each distribution. These quantities are calculated in a window of $2\sigma$ around the mean. The resolutions at eight simulated energies (from 1~GeV to 200~GeV) are then used to fit a curve of the form $\frac{a}{\sqrt{E}} \oplus c$ to obtain the stochastic and constant parameters for each absorber to scintillator proportion (electronic noise is not simulated in this context, so $b=0$). The resulting resolution curves and estimated parameters are shown in Figure \ref{fig:res_curve_prop} (left).

Using the parameters obtained, we can plug them back into Equation \ref{eq:enery_res} to obtain the resolution as a function of energy and absorber proportion. It is not possible to determine an ideal proportion that minimizes the resolution for all energies, as seen from the different minima in Figure \ref{fig:res_curve_prop}.

\begin{figure}[htb]
    \centering
    \includegraphics[width=0.85\linewidth]{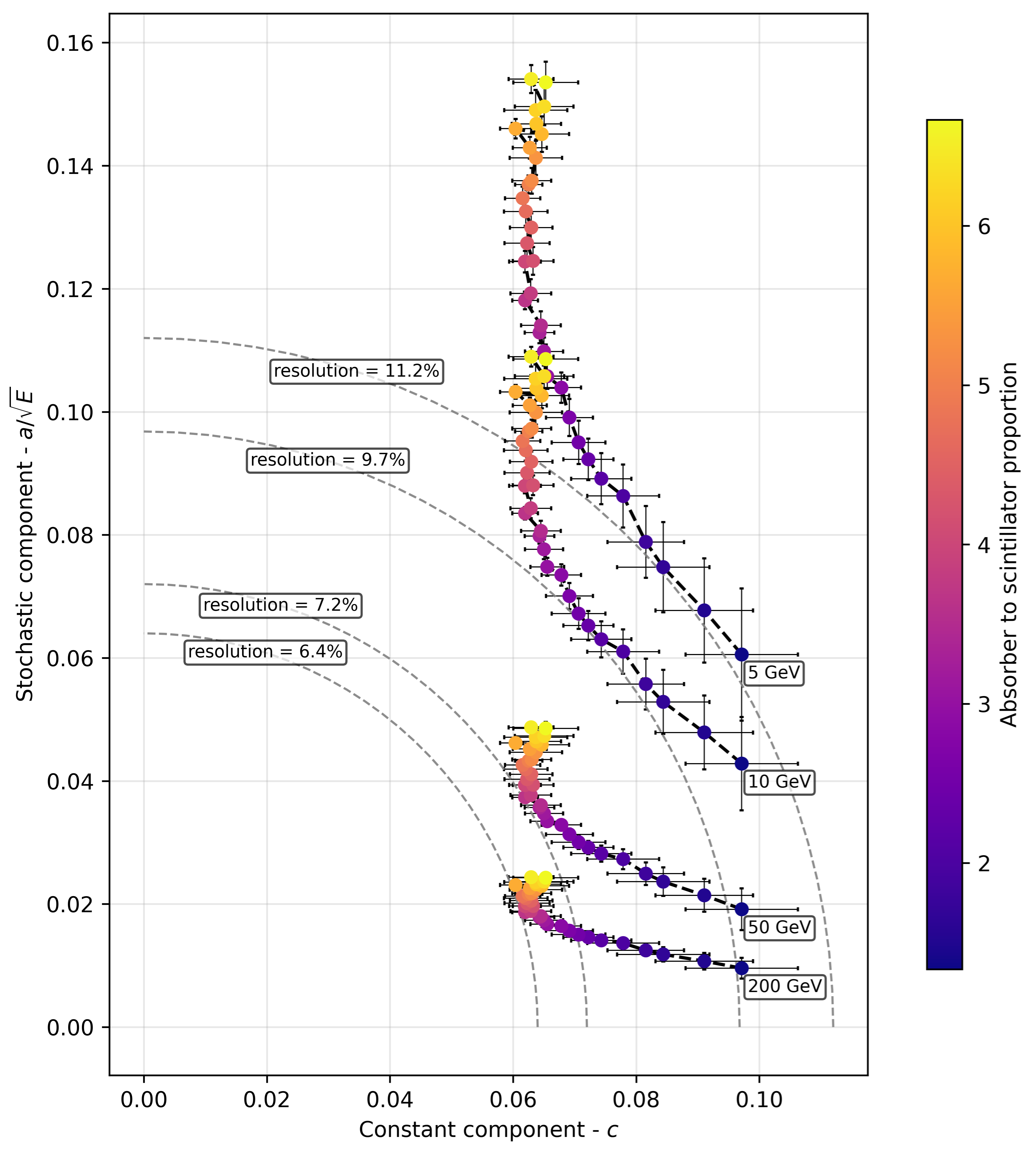}
    \caption{Stochastic and constant components of the energy resolution for different absorber proportions at four energies. The distance to the origin is the resolution, since the components are added in quadrature. The dashed lines are constant in resolution and are used to compare different proportions at a given energy. }
    \label{fig:component_plot}
\end{figure}

A different way to visualize the impact of the proportion on the energy resolution is to plot the contributions of the constant and stochastic components for a given energy. Since the components are added in quadrature, the distance to the origin represents the resolution. Figure \ref{fig:component_plot} shows how increasing the absorber proportion shifts the contribution from the constant component to the stochastic component. The ideal proportion for a given energy is the point closest to the origin. The $1/\sqrt{E}$ scaling of the stochastic component means that sampling becomes less relevant at high energies. Above 50~GeV, there are no further gains from using absorber plates thicker than 11~mm ($11/3$ proportion), as the improvement in the constant term stagnates for higher proportions. For energies below 50~GeV there is also some redundancy in the choice between the points among the equal resolution curve, albeit using smaller proportions than in the high energy limit. This shows that simply tuning the thickness of absorber material is not sufficient to achieve a dominating performance benefit.

\section{Conclusions and Outlook}

The optimal absorber to scintillator proportion increases with hadron energy. At energies above 50~GeV, using 11~mm thick iron plates in $11/3$ absorber to scintillator proportion was found to be the point of diminishing returns for energy resolution. Below 50~GeV, smaller proportions yield better resolutions due to the increased contribution of the stochastic component. There is no proportion $p$ that minimizes the resolution at all energies. Based on these results, a promising next step is to determine the longitudinal profile of deposited energy in the calorimeter and optimize the proportions $p_1 , p_2 , ... p_N$ for $N$ layers, targeting the proportion in each layer to accommodate the shower energy profile of typical hadrons and jets produced in the FCC-ee. A gradient-based optimization scheme, utilizing differentiable surrogate models of the detector response conditioned on the design parameters could provide crucial acceleration, since the variables involved here no longer allow a grid search approach.

\section*{Acknowledgments}

All authors are supported by project 2024.05140.CERN. I.O. is also supported by projects 2023.00042.RESTART and 2023.06052.CEECIND/CP2838/CT0001.

\end{document}